\documentclass{aa}

\usepackage{graphicx,natbib}
\usepackage{bm}
\usepackage{amsmath}

\usepackage{color}
\usepackage{ulem}

\begin{document}

   \title{Shock-reflected electrons and X-ray line spectra}

   \authorrunning{Dzif\v{c}\'akov\'a et al.}
   \titlerunning{Shock reflected electrons and X-ray line spectra}

   \author{E. Dzif\v{c}\'akov\'a, M. Vandas and M. Karlick\'y
          }
   \offprints{E. Dzif\v{c}\'akov\'a, \email{elena@asu.cas.cz}}

   \institute{Astronomical Institute of the Czech Academy of Sciences, Fri\v{c}ova 298,
251 65 Ond\v rejov, Czech Republic}
   \date{Received ; accepted }


  \abstract
   {}
 {  The aim of this paper is to try to explain the physical origin of the non-thermal electron distribution that is able to form the enhanced intensities of
 satellite lines in the X-ray line spectra observed during the impulsive phases of some solar flares.}
{Synthetic X-ray line spectra of the distributions composed of the distribution of shock reflected electrons and the background Maxwellian distribution are calculated in the approximation of non-Maxwellian ionization, recombination, excitation and de-excitation rates. The distribution of shock reflected electrons is determined analytically.}
 {We found that the distribution of electrons reflected at the nearly-perpendicular shock resembles, at its high-energy part, the so called $n$-distribution. Therefore it could be able to explain the enhanced intensities of \ion{Si}{xii}d satellite lines.  However, in the region immediately in front of the shock its effect is small  because electrons in background Maxwellian plasma are much more numerous there. Therefore, we propose a model in which the shock reflected electrons propagate to regions with smaller densities and different temperatures. Combining the distribution of the shock-reflected electrons with the Maxwellian distribution having different densities and temperatures we found that spectra with enhanced intensities of the satellite lines are formed at low densities and temperatures of the background plasma when the combined distribution is very similar to the $n$-distribution also in its low-energy part. In these cases, the distribution of the shock-reflected electrons controls the intensity ratio of the allowed \ion{Si}{xiii} and \ion{Si}{xiv} lines to the \ion{Si}{xii}d satellite lines. The high electron densities of the background plasma reduce the effect of shock-reflected electrons on the composed electron distribution function, which leads to the Maxwellian spectra. }
   {}


   \maketitle

\section{Introduction}
   It is  commonly known that solar-flare electron distributions can be strongly non-thermal. However, besides a high-energy tail, another kind of non-thermal distribution can also be present.  Analysis of several soft-X-ray-flare spectra recorded by the SOLFLEX Bragg crystal spectrometer on the P78-1 spacecraft  showed that the intensity ratios of \ion{Fe}{xxv} resonance line to \ion{Fe}{xxiv}d satellite lines could be explained by the presence of non-thermal energy distribution with a `bump' during the early and peak phases of the flare \citep{1987ApJ...319..541S}. They also concluded  that observed enhanced intensities of the satellite lines in comparison with Maxwellian synthetic spectra could not be assigned to the effect of a multithermal plasma.

High intensities of satellite lines are typical for the flare X-ray line spectra.  \cite{2006ApJ...638.1154P} used GOES flare temperatures to calculate intensities of satellite lines and compared them with those observed by the RESIK bent-crystal spectrometer aboard the CORONAS-F satellite \citep{2005SoPh..226...45S}.
RESIK spectra covered 3.3\,--\,6.1\,\AA\,and included emission lines of \ion{Si}, \ion{S}, \ion{Cl}, \ion{Ar}, \ion{K}{}, and many dielectronic satellite lines.
They showed that the observed intensities of satellite lines are typically two times higher than theoretical ones.

\begin{figure}
\centering
\includegraphics[width=8.3cm]{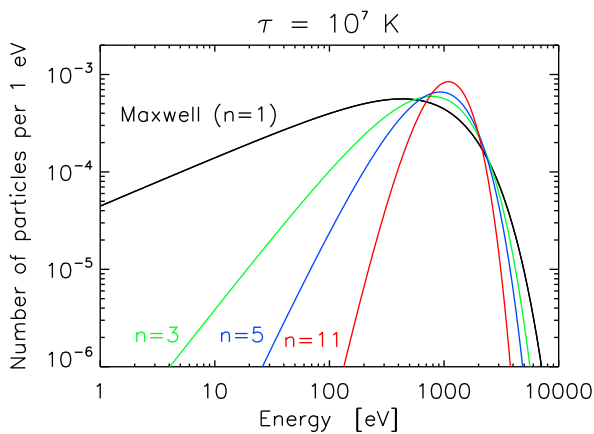}
\caption{Comparison of the Maxwellian distribution ({black line}) with $n$-distributions with  $n$\,=\,3 ({green}), 5 ({blue}), and 11 ({red}). The mean energy of distributions is the same. }
\label{n_dst}
\end{figure}

\cite{2008A&A...488..311D} adapted the analytic expression of peaked distribution proposed by  \cite{1987ApJ...319..541S} as $n$-distribution (Fig. \ref{n_dst})  to explain relative line intensities of silicon X-ray flare spectra from the RESIK. They also considered that line spectra could be multi-thermal and calculated their
differential emission measures (DEMs). These DEMs showed the presence of a dominant temperature component.  Comparison of synthetic spectra for derived DEM's with observations showed that they could not reproduce the intensity ratios of the allowed lines to the satellite ones  \citep[][Figs. 11 and 12]{2008A&A...488..311D}.  However, synthetic spectra for the $n$-distribution were able to describe relative intensities of both the allowed and satellite lines. Spectra corresponding to $n$-distribution were observed  during the impulsive phase of the flare. In decay phase and later, the spectra corresponded to an isothermal Maxwellian distribution \citep[][Fig. 13]{2008A&A...488..311D}.

Non-thermal components of the electron distribution during two flares using both RESIK and RHESSI spectra were analyzed by \cite{, 2011A&A...533A..81K}. They found that apart from a component corresponding to the electron beam, RHESSI spectra showed the other non-thermal component in keV energy range during the impulsive phase. This component was explained by the $n$-distribution with similar parameters to the $n$-distribution diagnosed independently from the silicon lines in the RESIK spectra. This non-thermal $n$-distribution was associated with the presence of the electron beam indicated by a radio burst.

\begin{figure}
\includegraphics[width=8.3cm]{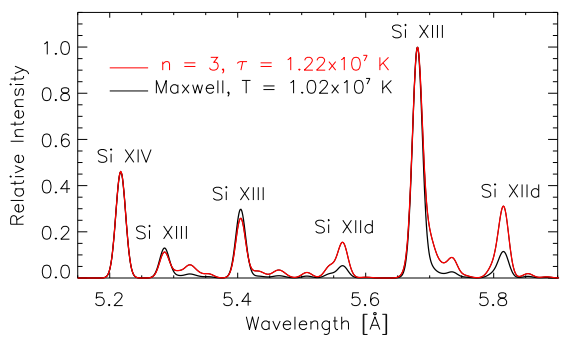}
\caption{X-ray line spectra for the Maxwellian distribution ({black line})  and the $n$-distribution with  $n$\,=\,3 ({red}). The intensities of the satellite lines \ion{Si}{xii}d are enhanced for the $n$-distribution in the comparison with the Maxwellian spectrum. The line ratio of \ion{Si}{xiv} 5.22 \AA\, to \ion{Si}{xiii} 5.68 \AA\,line is the same in both spectra.
}
\label{sp_n_k}
\end{figure}

Fig. \ref{sp_n_k} demonstrates the effect of the $n$-distribution on the relative line intensities. The synthetic spectra for the Maxwellian distribution and for the $n$-distribution have the same intensity ratio of allowed lines,  \ion{Si}{xiv} 5.22\,\AA\,to \ion{Si}{xiii} 5.68\,\AA. Similarly to observations, the intensities of the satellite lines \ion{Si}{xii}d\,5.82\,\AA\,and 5.56\,\AA\,are more than two times higher for the $n$-distribution than for the Maxwellian one.

The non-Maxwellian  $n$-distribution is usually expressed as \citep[e.g.,][]{1998SoPh..178..317D}
\begin{equation}\label{eq1}
f_n(E) = B_n \frac{2E^{1/2}}{(k_BT)^{3/2}\sqrt{\pi}}
\left(\frac{E}{k_BT}
\right)^\frac{n-1}{2} \exp\left[-\frac{E}{k_BT}\right],
\end{equation}
\begin{equation}
B_n = \frac{\sqrt{\pi}}{2\Gamma(n/2+1)},
\end{equation}
where $E$ is the electron energy, $k_B$ the Boltzmann constant, and $n \in \left<1,\infty\right)$ the parameter of the distribution. The distribution is normalized to unity. For $n$ = 1, the distribution is Maxwellian and then $T$ is its temperature. The mean energy of the $n$-distribution depends on $n$
\begin{equation}
\langle E \rangle  = \left(\frac{n}{2}+1\right) k_B T = \frac{3}{2} k_B \tau,
\end{equation}
where $\tau$ is the pseudo-temperature and $T$ is just a parameter of the distribution.
Contrary to the Maxwellian distribution, the $n$-distribution  has narrower peak and is steeper at its high-energy part when their mean energies are the same (Fig. \ref{n_dst}).

Up to now this electron distribution was used only as the parametric one and its physical origin remains unclear.
There have been several attempts to explain processes forming the $n$-distribution in solar flare conditions. Some of its aspects, especially its high-energy part, were explained considering the return current in the beam-plasma system \citep{2008SoPh..250..329D,2012A&A...537A..36K} or as the distribution of electrons accelerated in the double-layer \citep{2012ApJ...750...49K}. However, both explanations have some drawbacks. While in the return-current explanation the drift associated with the return current needs to be greater that the thermal velocity of the background plasma, in the case of the double-layer explanation, an existence of the double layer is not still confirmed.

Recently, \cite{2016A&A...591A.127V} studied the distributions of electrons reflected at the nearly perpendicular shock. They found that these distributions exhibit some aspects of the $n$-distribution.  However, in the close vicinity of the shock, the electrons in this distribution are much less numerous than that of the background plasma (which serves electrons for reflection in the shock. Thus, an effect of the $n$-distribution on the X-ray line spectra from regions close to the shock is nearly negligible.

Therefore, in the present paper we assume that the shock-reflected electrons propagate some distance from the shock where the background (Maxwellian) plasma electrons could have smaller density and different temperature compared with those at the shock vicinity, and thus produce some effects on the resulting X-ray line spectra.

This paper is organized as follows: In Section 2 we present distributions of electrons reflected at the shocks having different parameters. Then in Section
3 we describe details of calculations of the X-ray line spectra. The resulting spectra for various distributions of the shock-reflected electrons as well as
for various distributions of the background plasma are presented in Section 4.
Finally, a discussion and conclusions are presented in Sections 5 and 6, respectively.

\section{Distribution of electrons reflected at a nearly perpendicular shock}

\begin{figure}
\centering
\includegraphics[width=7.5cm]{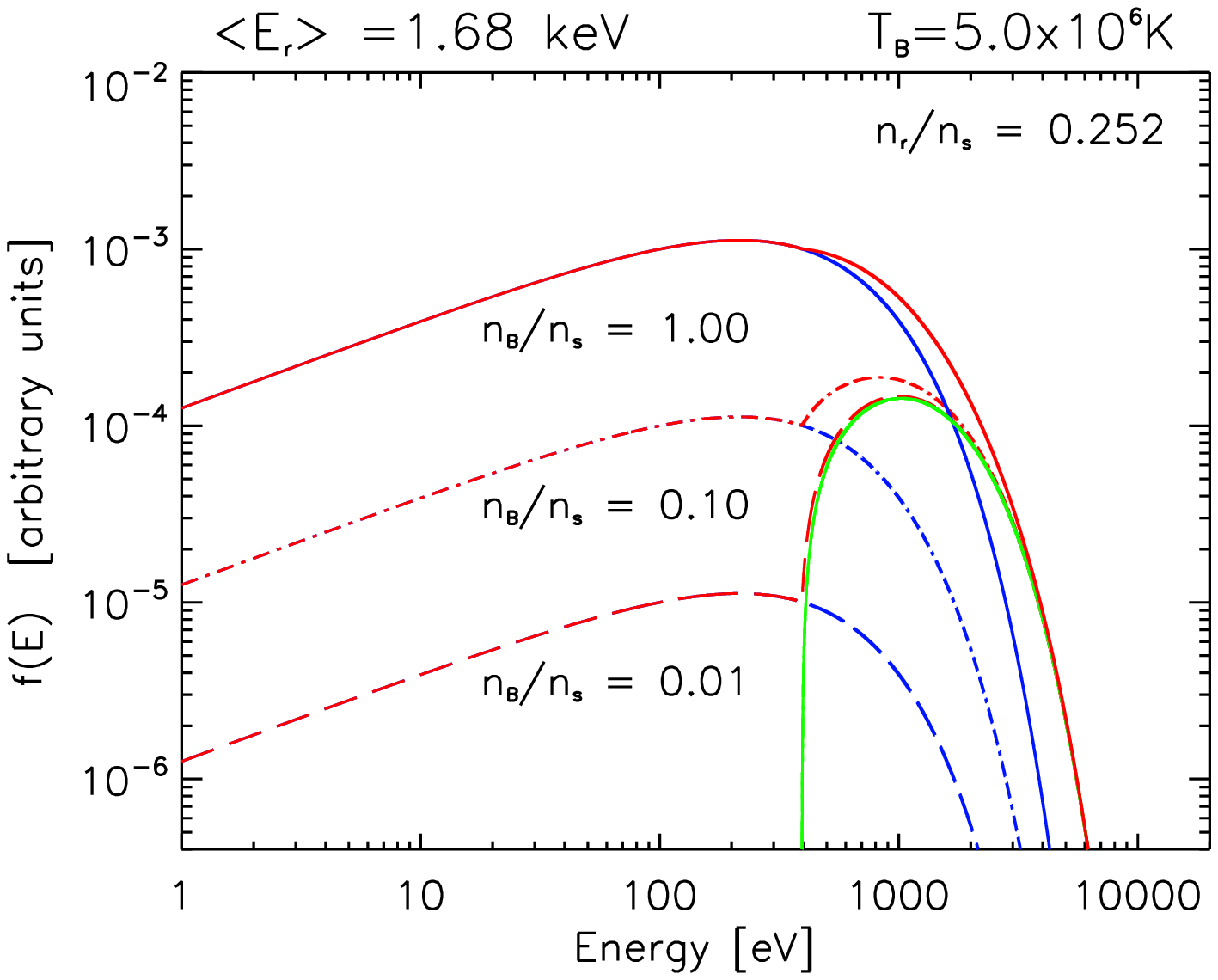}
\includegraphics[width=7.5cm]{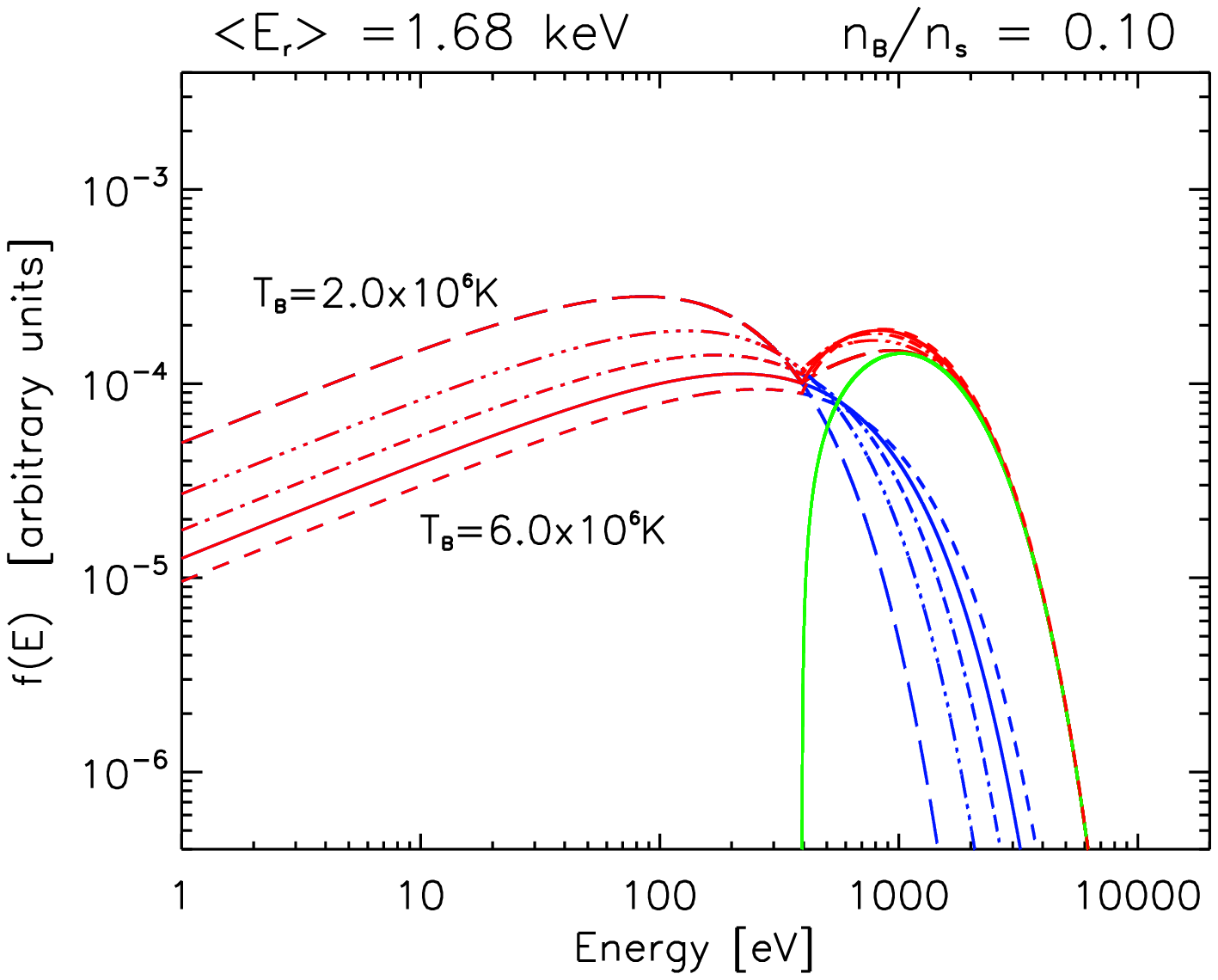}
\includegraphics[width=7.5cm]{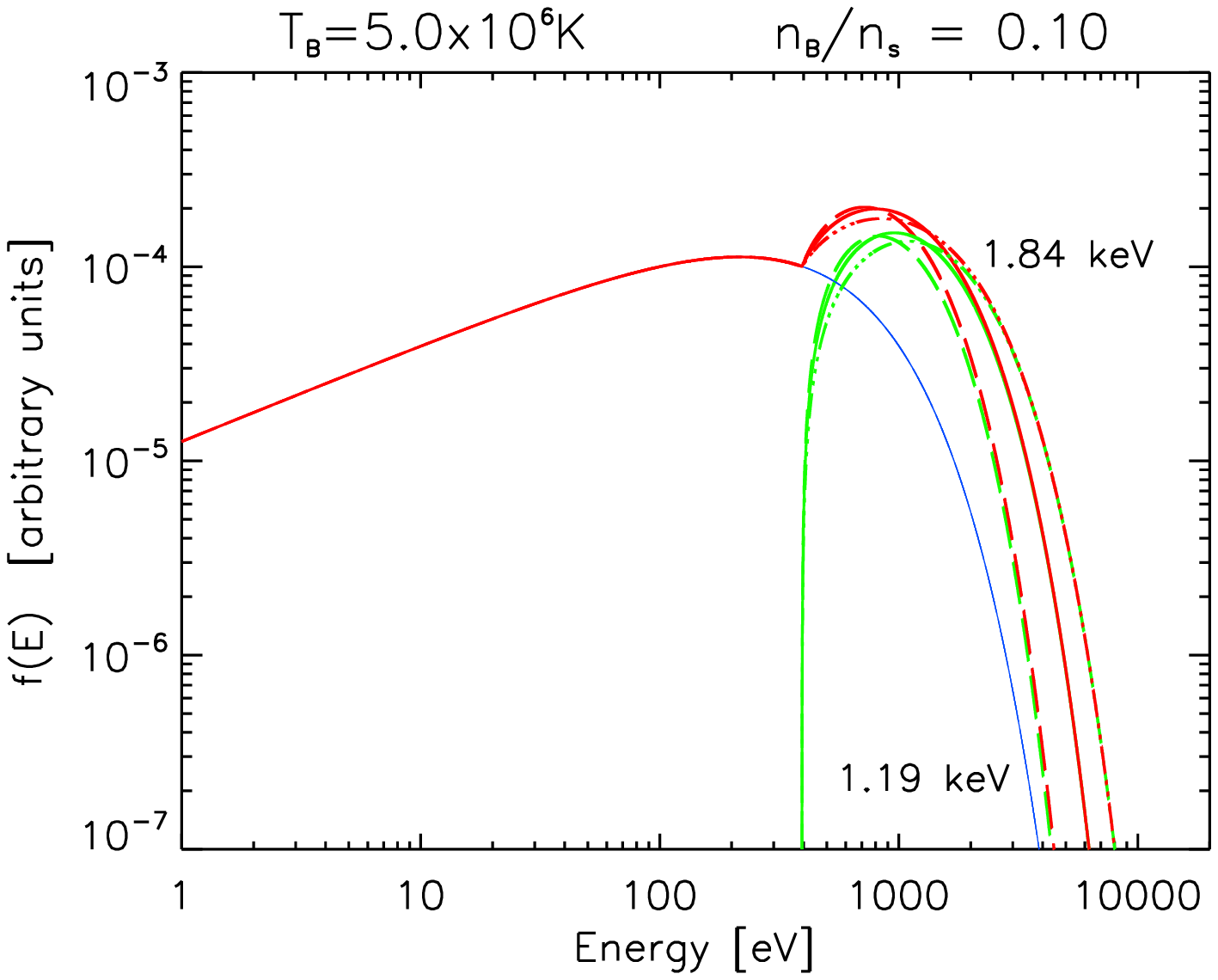}
\caption{Electron distributions ({red lines}) composed from the shock reflected electrons ({green lines})
with different background Maxwellian distribution ({blue lines}).
{Top:} The distributions composed from the distribution of reflected electrons
with $\langle E_r\rangle$\,=\,1.68~keV and $T_s$~=~6~MK (case~4, Tab.~\ref{tab1}) with Maxwellian distributions having different
electron densities $n_B$\,=\,0.01, 0.1, and 1.0 and fixed temperature $T_B$\,=\,5~MK.
{Middle:} The distributions composed from the distribution of reflected electrons
(case~4) with the Maxwellian distributions having $n_B/n_s$\,=\,0.1
and temperature $T_B$\,=\,2, 3, 4, 5, and 6~MK.
{Bottom:} The distributions composed from different distributions of reflected electrons ($\langle E_r\rangle$\,=\,1.19, 1.52, and 1.84~keV)
(case~1, 3, and 5) with the Maxwellian distribution with
$T_B$\,=\,5~MK and $n_B/n_s$\,=\,0.1.}
\label{Fig:distr}
\end{figure}

Since the 1970s, through in-situ spacecraft observations, it has been known that  nearly perpendicular, fast, collisionless shocks are capable of accelerating electrons. \cite{W84} and \cite{LM84} suggested that the  nearly perpendicular shock acts as a fast moving magnetic mirror. This kinematic point of view was subsequently supplemented by a dynamic one \citep{KW89,V89b}. Electrons drift in the shock layer due to gradient and curvature drifts; this drift is against the induced electric field ($\mathbf{V} \times \mathbf{B}$), so electrons gain energy. Some of them are reflected at the magnetic field increase and form a beam traveling upstream away from the shock along magnetic field lines.

For a plane shock wave and thermal plasma the distribution function of reflected electrons can be expressed analytically \citep{2016A&A...591A.127V}
\begin{eqnarray}
f_r(E) & = & \frac{n_s}{4 \sqrt{\pi E_B E_s}} \nonumber \\
&  & \times
\left[\mathrm{e}^{\frac{4 E_B}{E_s} \cos \theta_c
\left(\sqrt{\frac{E}{E_B}-\sin^2 \theta_c-\frac{e \Delta \Phi}{E_B}
\tan^2 \theta_c}-\cos \theta_c \right)} \right. \nonumber \\
& & \left. -1\right]
\mathrm{e}^{-\frac{E}{E_s}}  \hspace*{1.5cm}  E > E_m  ,  \label{frmp} \\
     &  = &  0 \hspace*{2.75cm} E \le E_m  , \nonumber
\end{eqnarray}
\begin{equation} \label{Em}
E_m  =  E_B + e \Delta \Phi \tan^2 \theta_c  ,
\end{equation}
where $e$ is the elementary charge, $\Delta \Phi$ is the electrostatic cross-shock potential, $\sin \theta_c = \sqrt{B_1/B_2}$, $B_1$ and $B_2$ are the upstream and downstream magnetic fields, respectively, $E_B=\frac{1}{2} m_e V_B^2$, $V_B=V_{1n}/\cos \theta_{Bn}$, $V_{1n}$ is the normal component of the upstream plasma bulk velocity with respect to the shock front, $m_e$ is the electron mass, and $\theta_{Bn}$ is the angle between the shock normal and the upstream magnetic field. The energy $E_s=\frac{1}{2} m_e v_s^2 =k_B T_s$ because the upstream plasma is assumed to be thermal with electron density $n_s$, temperature $T_s$, and thermal velocity $v_s$ (`$s$' for `seed' because these electrons serve as seed particles for the acceleration process), so the initial distribution function of electrons is $n_s f_1(E)$ where $f_1$ is from Eq.~(\ref{eq1}) with $n=1$ and $T=T_s$. We note that $n$ with a subscript means a number density throughout the paper, while the sole $n$ is the parameter of the $n$-distribution function. The distribution function~(\ref{frmp}) is not normalized to unity but is related to the number density $n_s$ of seed particles.  The velocity $V_B$ is the velocity of the magnetic mirror; it must be comparable to electron velocities for an efficient electron acceleration. Because the upstream plasma velocity $V_1$ is much smaller than the thermal speed $v_s$, $\theta_{Bn}$ must be close to $90^\circ$ (i.e.,   a nearly perpendicular shock).

The number density $n_r$ and mean energy $\langle E_r \rangle$ of reflected electrons are
\begin{eqnarray}
n_r & = & \int\limits_{E_m}^\infty f_r(E) \, \mathrm{d}E \nonumber \\
 & = & \frac{n_s}{2} \cos \theta_c
\, \left[1+\mathrm{erf} \left(\cos \theta_c \sqrt{\frac{E_B}{E_s}} \right)
\right] \nonumber \\
& & \times \mathrm{e}^{-\frac{E_B \sin^2 \theta_c+
e \Delta \Phi \tan^2 \theta_c}{E_s}} ,  \label{nr} \\
\langle E_r \rangle & = & \frac{1}{n_r} \int\limits_{E_m}^\infty E f_r(E) \,
\mathrm{d}E \nonumber \\
& = & \frac{3}{2} E_s+E_B (1+3 \cos^2 \theta_c)
+e \Delta \Phi \tan^2 \theta_c  \nonumber \\
& & + 3 \cos \theta_c \sqrt{\frac{E_B E_s}{\pi}}
\, \left[1+\mathrm{erf} \left(\cos \theta_c \sqrt{\frac{E_B}{E_s}} \right)
\right]^{-1} \nonumber \\
& & \times \mathrm{e}^{-\frac{E_B}{E_s} \cos^2 \theta_c} , \label{Er}
\end{eqnarray}
where $\mathrm{erf}$ is the error function.

\begin{figure}
\centering
\includegraphics[width=8.0cm]{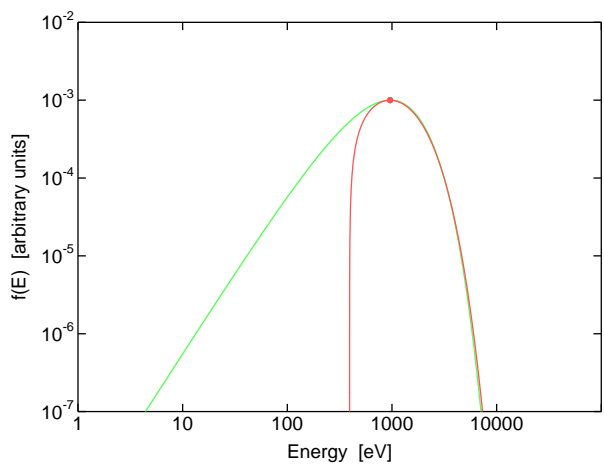}
\caption{Result of the fitting procedure for case~3. The distribution function of reflected electrons is given in red, and a corresponding $n$-distribution function in green. The bullet points to the common maximum of the distribution functions. }
\label{figfit}
\end{figure}

\begin{table}
\caption{The list of model parameters, characterization of the corresponding distribution function of reflected electrons and its relationship to the $n$-distribution function}
\begin{tabular}[t]{rcccrl}
\hline
Case & $n_r/n_s$ & $\langle E_r \rangle$ & $E_\mathrm{max}$ &
\multicolumn{1}{c}{$n$} & Parameter \\
&  & [keV] & [keV] &  & change   \\
\hline
 1 & 0.157 & 1.188 & 0.820 & 6.2 & $T_s$~=~3~MK \\
 2 & 0.201 & 1.354 & 0.890 & 4.9 & $T_s$~=~4~MK  \\
 3 & 0.231 & 1.517 & 0.959 & 4.2 \\
 4 & 0.252 & 1.677 & 1.027 & 3.7 & $T_s$~=~6~MK  \\
 5 & 0.268 & 1.835 & 1.093 & 3.4 & $T_s$~=~7~MK  \\
 6 & 0.315 & 1.384 & 0.826 & 3.3   & $\Delta \Phi = 0$ \\
 7 & 0.339 & 1.566 & 0.979 & 3.7   & $B_2/B_1=2$ \\
 8 & 0.431 & 1.618 & 1.010 & 3.5    & $B_2/B_1=2.5$ \\
 9 & 0.162 & 2.207 & 1.553 & 6.6    & $\theta_{Bn}=86^\circ$ \\
10 & 0.091 & 3.139 & 2.412 & 10.5  & $\theta_{Bn}=87^\circ$ \\
\hline
\end{tabular}
\label{tab1}
\end{table}

For the numerical results presented in the following, we started with the same shock parameters to those used in \cite{2016A&A...591A.127V}: $V_{1n}$~=~1000~km/s, $T_s$~=~5~MK, $B_2/B_1=1.6$, $\theta_{Bn}$~=~84$^\circ$, and $\Delta \Phi$~=~80~V.   The shock has Mach number 1.5, which is consistent with the value recently reported from observations \citep{2015Sci...350.1238C}. The given parameters correspond to case 3 listed in Tab.~\ref{tab1}. For other considered cases, only one parameter was varied as shown in the last column of the table. Cases 1--5 were used for X-ray spectra computations; they represent a variation in $T_s$, the seed plasma temperature (the temperature upstream of the shock). Reflected electrons from these cases were combined with background Maxwellian plasma with the temperature $T_B$ (from 2 to 6~MK) and density $n_B$ ($n_B/n_s$ from 0.01 to 1). Construction of the resulting distribution functions is shown in Fig.~\ref{Fig:distr}.

Table~\ref{tab1} lists the relative number density $n_r/n_s$ of reflected electrons, their mean energy $\langle E_r \rangle$  and the energy $E_\mathrm{max}$ where the distribution function reaches a maximum. The fifth column contains $n$, the parameter of an $n$-distribution function that best fits the distribution function of reflected electrons in its high-energy part. The result of the fit for case~3 is shown in Fig.~\ref{figfit}.

The fitting procedure was the following: A value of $n$ is determined that minimizes the difference between the distribution functions (of
reflected electrons and the $n$-distribution) in logarithmic scale within the energy interval from $E_\mathrm{max}$ to an energy where the distribution function of reflected electrons fell by four orders of magnitude. $E_\mathrm{max}$ is the energy where the distribution function of reflected electrons has maximum (it is given in Tab.~ref{tab1}). The $n$-distribution function also depends on the parameter $T$. It is calculated for a given $n$ from the formula $E_\mathrm{max}=\frac{1}{2} n k_B T$ so that the maxima of both distribution functions coincided (see Fig.~\ref{figfit})

From cases~1--5 we see that the value of $n$ decreases with increasing $T_s$. In order to gain insight into how $n$ changes with variations of shock parameters, additional cases are given in Tab.~\ref{tab1}. Case~6 was treated in \cite{2016A&A...591A.127V} where we tried to find shock parameters to fit the $n$-distribution function with $n=3$ in its high-energy part. This was done by trial and error and visual inspection. Here a more objective procedure yields $n=3.3$, not far from the desired value of~3. We see that the value of $n$ decreases with increasing jump in magnetic field at the shock wave, or with decreasing electrostatic potential. It increases with $\theta_{Bn}$ rise, and this increase is very rapid above $87^\circ$.

\section{Calculation of synthetic X-ray line spectra}

\begin{figure}[h]
\centering
\includegraphics[width=8.5cm]{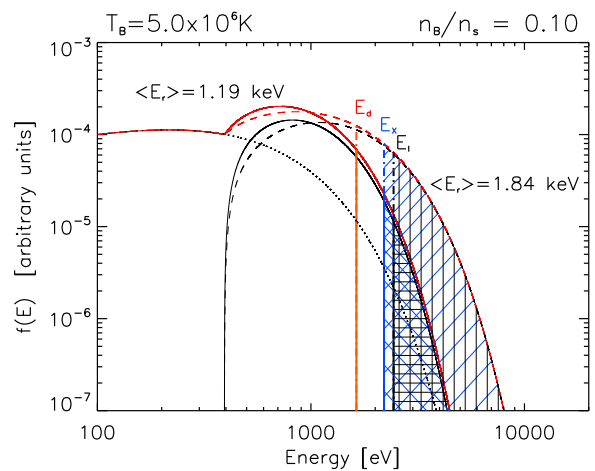}
\caption{Two composed electron distributions for $T_B$=5$\times10^6$ K, $n_B/n_s$=0.10, and $\langle E_r \rangle$=1.19 keV and 1.84 keV  with marked excitation energy $E_d$ of the satellite line \ion{Si}{xii}d 5.82 \AA, excitation energy $E_x$ of the allowed line \ion{Si}{xiii} 5.68 \AA,\, and the relative number of electrons over distribution that contributed to the excitation rate (crossway hatched areas),  the ionization energy $E_I$ of \ion{Si}{xiii} to \ion{Si}{xiv} and relative number of electrons over distribution that contributed to the ionization rate (horizontally and vertically hatched areas).}
\label{satellite}
\end{figure}

We calculated X-ray line synthetic spectra for silicon in the same spectral interval as in \cite{2008A&A...488..311D}, that is, the RESIK spectral region 5.2\,--\,6.0\,\AA,\,in order to be able to compare them with observations.  RESIK was a high-resolution crystal X-ray spectrometer  \citep{2005SoPh..226...45S} on board the Russian CORONAS-F mission.  The RESIK spectral region is very useful for the diagnostics of electron distribution because it contains both allowed and satellite lines.  In contrast to allowed lines,  whose excitation rate is an integral of the product of the cross-section with distribution function from the excitation energy $E_x$ to infinity (Fig.~\ref{satellite}, crossway hatched areas), the intensities of the satellite lines depend on the number of electrons in the distribution with the energy corresponding to the excitation energy of the double-excited state of ion $E_d$ (Fig.~\ref{satellite}). This means that satellite lines are able to sample the electron distribution and their atypical non-Maxwellian intensities can be very sensitive indicators of the presence of a non-thermal electron distribution  \citep{1979MNRAS.189..319G}. The intensities of \ion{Si}{xii}d satellite lines also depend on the number of the recombining \ion{Si}{xiii} ions, therefore their ratios to \ion{Si}{xiii} allowed lines do not depend on the ionization state of plasma.

The synthetic silicon X-ray line spectra for the $n$-distributions were calculated by \cite{2008A&A...488..311D}. The enhanced intensities of the satellite lines \ion{Si}{xii}d for $n$-distribution in comparison with Maxwellian intensities mean that the electron distribution must have an enhanced number of electrons with the energy  corresponding to their excitation energies in comparison with the number of electrons in the Maxwellian distribution. However, the behavior of the whole spectrum with changes in temperature and distribution function is not so simple. The ratio of \ion{Si}{xiv} 5.22 \AA\,to \ion{Si}{xiii} 5.68\,\AA \,also depends on the electron distribution through both the ionization and excitation state of the plasma. This ratio increases with temperature or pseudo-temperature; differently, however, for different distributions.
On the contrary, the relative intensity of \ion{Si}{xiii}d\,5.82\,\AA\,to\,\ion{Si}{xiii}\,5.68\,\AA\,decreases with the temperature.

Although the distribution functions of shock-reflected electrons are similar to $n$-distribution, a background Maxwellian distribution is also present and the true distribution is a combination of both distributions. Therefore, we performed new calculations of the ionization and excitation state for particular composed distributions to determine the plasma line emission. We suppose an equilibrium state in our calculation because high electron densities of flaring plasma lead to equilibrium times for the ionization around or below 1 sec. Equilibrium times for the electron excitation are even shorter than for the ionization \citep[e.g.,][]{2009A&A...502..409B}.  The line emissivity, $\varepsilon_{ij}$, can be written
\begin{equation}
 \varepsilon_{ij}=\frac{hc}{\lambda_{ij}} A_{ij} N^{+k}_{\mathrm{X},i}=\frac{hc}{\lambda_{ij}} A_{ij} A_{\mathrm{X}}  \frac{N^{+k}_{\mathrm{X},i}}{
N^{+k}_{\mathrm{X}}} \frac{N^{+k}_{X}}{N_{\mathrm{X}}}\frac{N_{\mathrm{H}}}{N_{\mathrm{e}}} N_{\mathrm{e}}
,\end{equation}
where $h$ is the Planck constant, $c$ is the speed of light, $\lambda_{ij}$ is the wavelength of the transition between atomic level $i$ and $j$, $A_{ij}$ is the Einstein coefficient for spontaneous emission, $N^{+k}_{\mathrm{X},i}$ is the number of ions with the excited level $i$, $N^{+k}_{\mathrm{X}}/N_{\mathrm{X}}$ is the abundance of $+k$-times ionized ions relative to the total number of ions for element $\mathrm{X}$, $A_{\mathrm{X}}$ is the abundance of the element $\mathrm{X}$  relative to hydrogen, $N_{\mathrm{H}}$ is the total number of hydrogen ions, and $N_{\mathrm{e}}$ is the electron density.  The line intensity is simply an integral of the emissivity along the line of sight in the optically thin coronal plasma.

In equilibrium state, the excitation equilibrium determines the ratio $N^{+k}_{\mathrm{X},i} / N^{+k}_{\mathrm{X}}$ and the ionization equilibrium  gives $N^{+k}_{\mathrm{X}}/N_{\mathrm{X}}$.

For the ionization, the direct electron ionization and autoionization are important in the coronal conditions. The dominant recombination processes are the radiative recombination and dielectronic recombination. In the equilibrium state, the total ionization is compensated by the total recombination.

The rate $R$ of any elementary process can be written:
\begin{equation}
    R=\langle\sigma v\rangle=\int_{0}^{\infty} \sigma f(E)~\left(\frac{2{E}}{m}\right)^{1/2} \mathrm{d} E
    \label{Eq:rates}
,\end{equation}
where $\sigma$ is the cross-section for the specific elementary process, $v$ is the electron velocity, $E$ is the electron energy, and $f(E)$ is the electron
distribution as a  function of the energy. The ionization rates for any electron distribution can be directly calculated from the ionization cross-sections based on  the atomic data of \citet{Dere07} and available in  the CHIANTI  8.0 database \citep{DelZanna15}.

The method of \citet{Dzifcakova92}, \citet{Wannawichian03}, and \citet{Dzifcakova13} has been used to determine radiative recombination rates. We assumed that the cross-section for the radiative recombination, $\sigma_\mathrm{RR}$, can be expressed \citep[e.g.,][]{Osterbrock74}
\begin{equation}
\sigma_{\mathrm{RR}}=C_{\mathrm{RR}}/{E}^{\eta+0.5}
,\end{equation}
where $C_{\mathrm{RR}}$ is a constant and $\eta+0.5$ is a power-law index. The parameters $C_{\mathrm{RR}}$ and $\eta$ were taken from  \citet{Aldrovandi73}, \citet{Landini90}, \citet{Shull82}, \citet{Mazzotta98}, and \citet{Badnell06}.

For the dielectronic recombination rate, the following expression, valid for any distribution function, has been used  \citep{Dzifcakova92,Dzifcakova13}
\begin{equation}
R_{\mathrm{DR}}=\sum_m A_m \frac{\pi ^{1/2}k_B^{3/2}}{2} \frac{f({ E}_m) }{{ E}_m^{1/2}}
,\end{equation}
where $A_m$ and ${ E}_m$ are the parameters from the Maxwellian approximations of the dielectronic recombination rates. The parameters have been taken from
\citet{2012A&A...537A..40A},  \citet{2006A&A...447.1165A}, \citet{2007A&A474.1051A}, \citet{2004A&A...426..699Z}, \citet{2006A&A...447..379Z}, \citet{2005A&A...438..743Z}, \citet{2004A&A...425.1153M}, \citet{2005A&A...440.1203Z}, \citet{2004A&A...420..775A}, \citet{2003A&A...412..597C}, \citet{2004A&A...417.1183C}, \citet{2007A&A...466..755B}, and \citet{2006A&A...447..389B} and can be found in the CHIANTI database \citep{Dere09} .

For the excitation equilibrium,  the electron collisional excitation and deexcitation together with the spontaneous radiative decay transitions are important. The collision strengths $\Omega_{ij}$ instead of the cross-sections $\sigma_{ij}$ are commonly used for calculation of the electron excitation rates:
\begin{equation}
    \sigma_{ij} = \frac{\Omega_{ij}}{ \omega_{i} E_{i}}\pi a_{0}^{2}\,,
    \label{Eq:Omega}
\end{equation}
where $\omega_{i}$ is the statistical weight of the level $i$, $E_{i}$ is the incident electron energy,  and $a_{0}$ is the Bohr radius.

The spline approximations of the Maxwellian-averaged collision strengths for the majority of the astronomically interesting ions of elements from \ion{H}{} to \ion{Zn}{} can be found in the CHIANTI database \citep{Dere97,DelZanna15}. The original $\Omega$s are usually inaccessible because of their huge data volumes. Therefore,  \citet{Dzifcakova15} developed method to calculate $\Omega$ for approximation of the non-Maxwellian excitation and de-excitation rates. The tests of this method on the $\kappa$-distributions for the KAPPA package  \citep{Dzifcakova15} showed precision within 5--10\% in comparison with direct numerical calculations. The $\Omega$ approximations for \ion{Si}{xii} - \ion{Si}{xiv} based on the atomic data of \citet{1990ADNDT..44...31Z},  \citet{2006ApJ...638.1154P}, \citet{1978ADNDT..21...49V}, \citet{1980ADNDT..25..311V}, \citet{1983ADNDTS}, \citet{1987ApJS...63..487Z}, \citet{Aggarwal1992} and contained in the CHIANTI database were used to calculate the X-ray spectra in the RESIK 5.2\,-\,6.0 \AA\, spectral window.

\begin{figure}
\centering
\includegraphics[width=8.4cm]{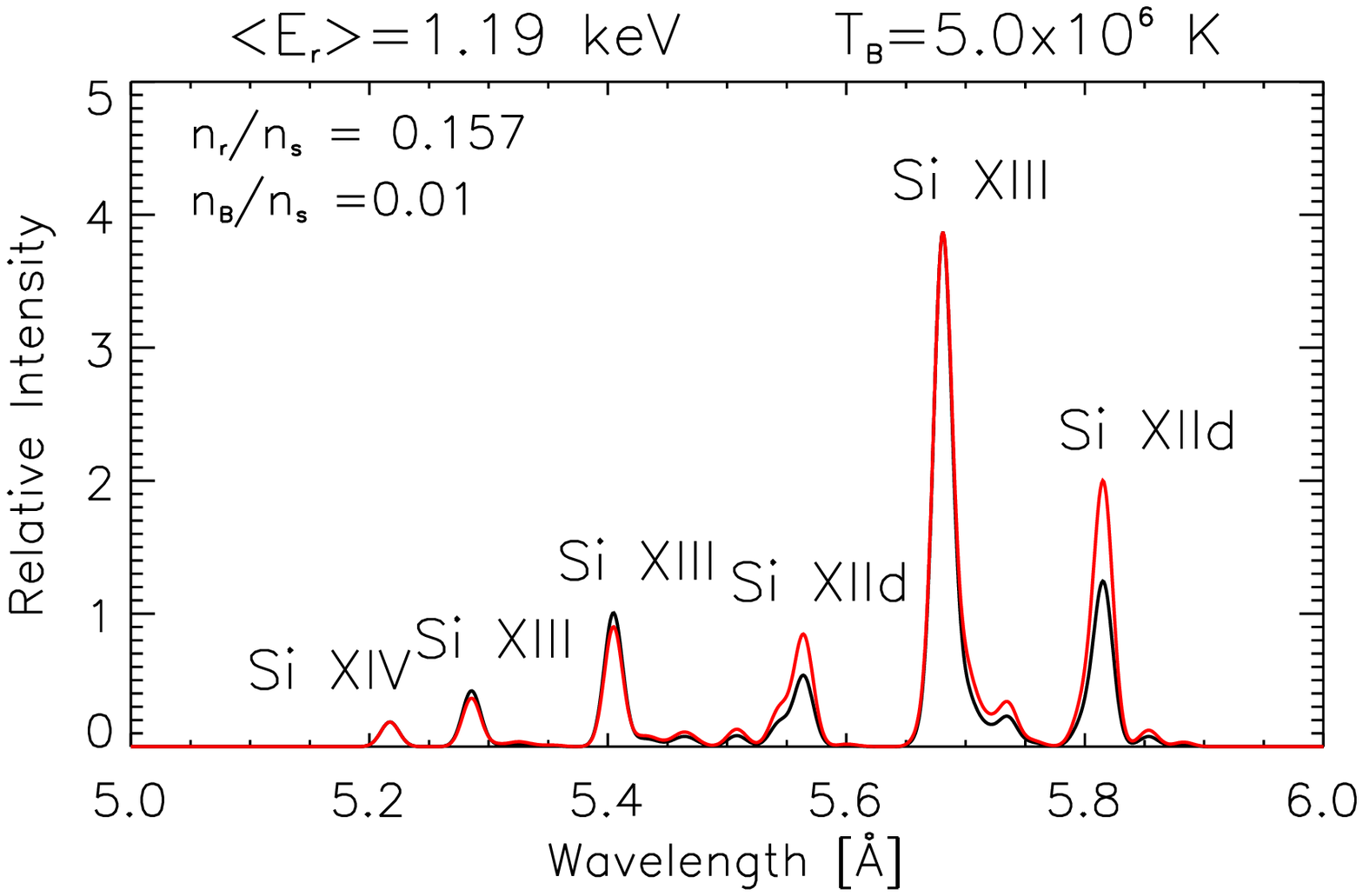}
\includegraphics[width=8.4cm]{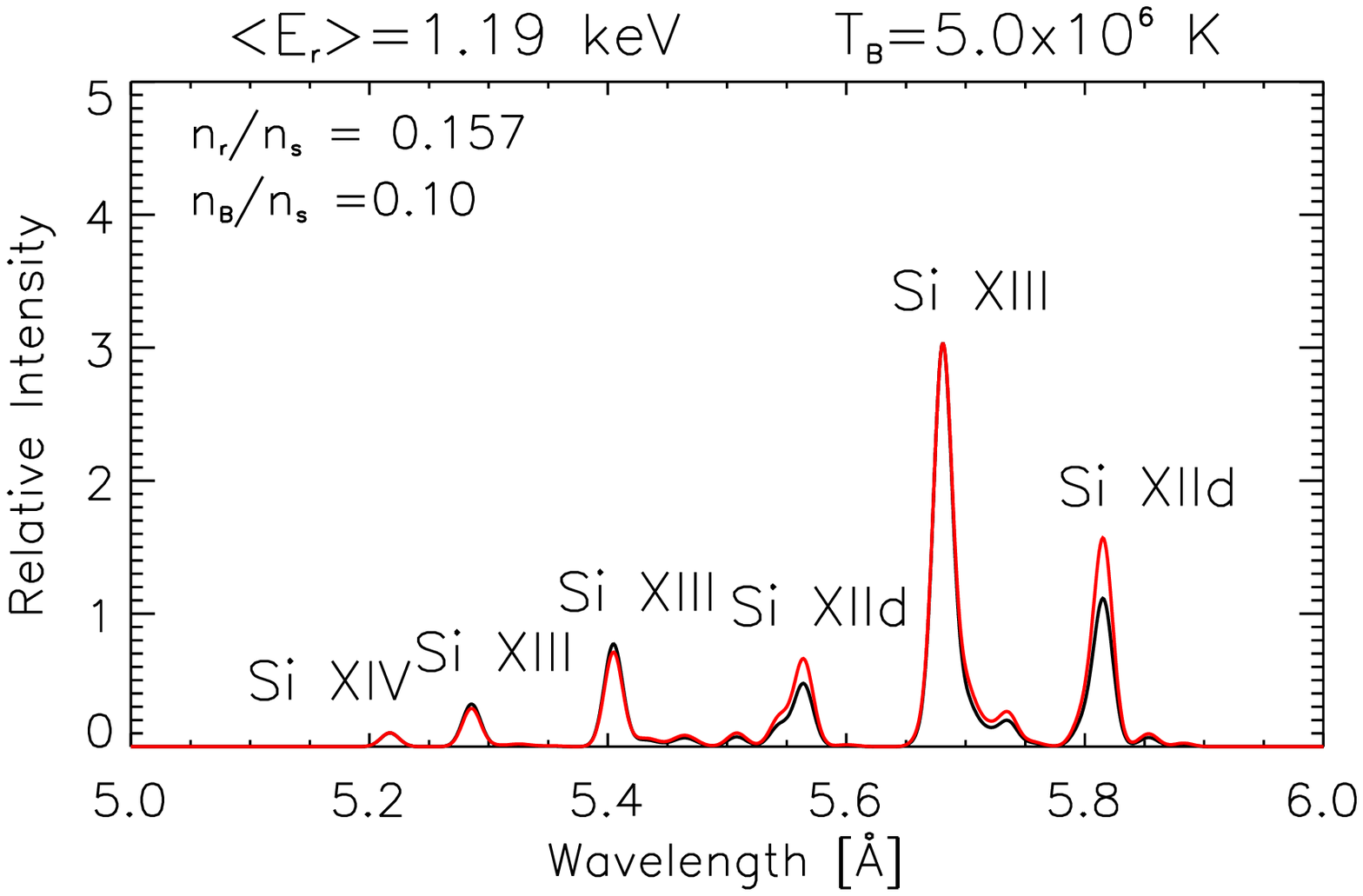}
\includegraphics[width=8.4cm]{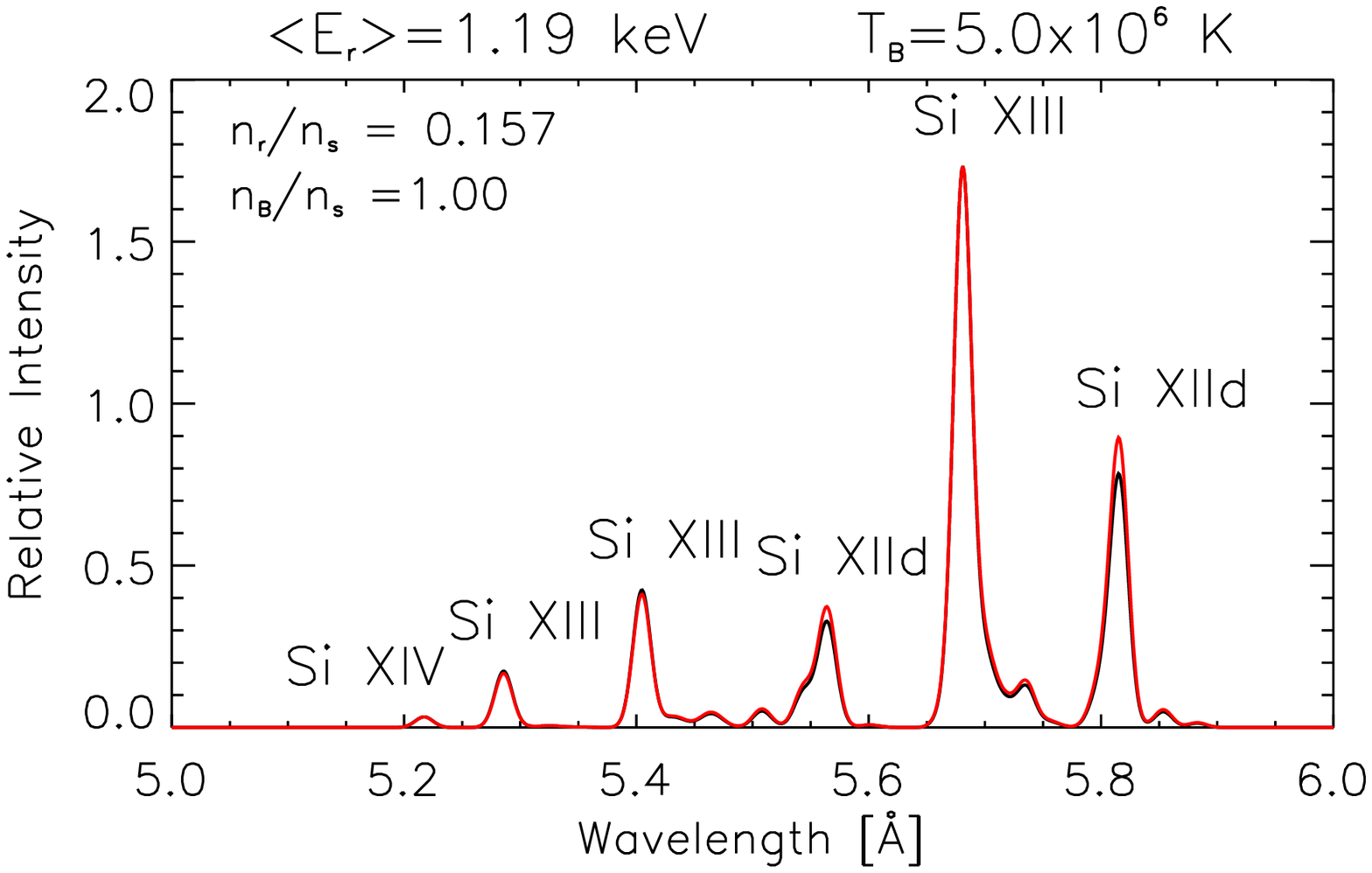}
\caption{Synthetic spectra of the composed distributions (red lines)
in dependence on the density
$n_B$ of the background plasma (from top to bottom). For comparison, the spectra of Maxwellian distributions
are added (black lines).}
\label{sp_comp1}
\end{figure}

\begin{figure}
\centering
\includegraphics[width=8.4cm]{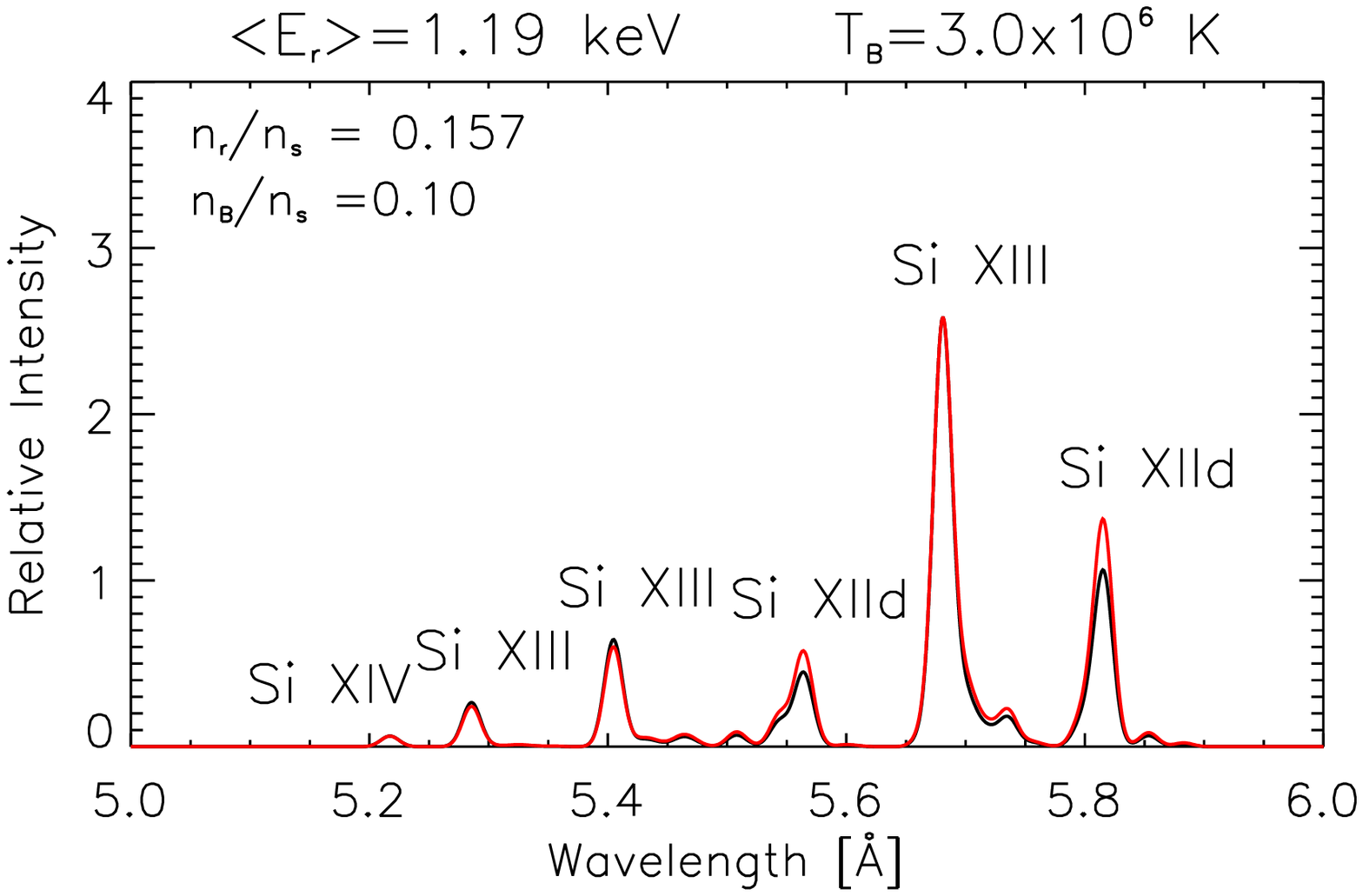}
\includegraphics[width=8.4cm]{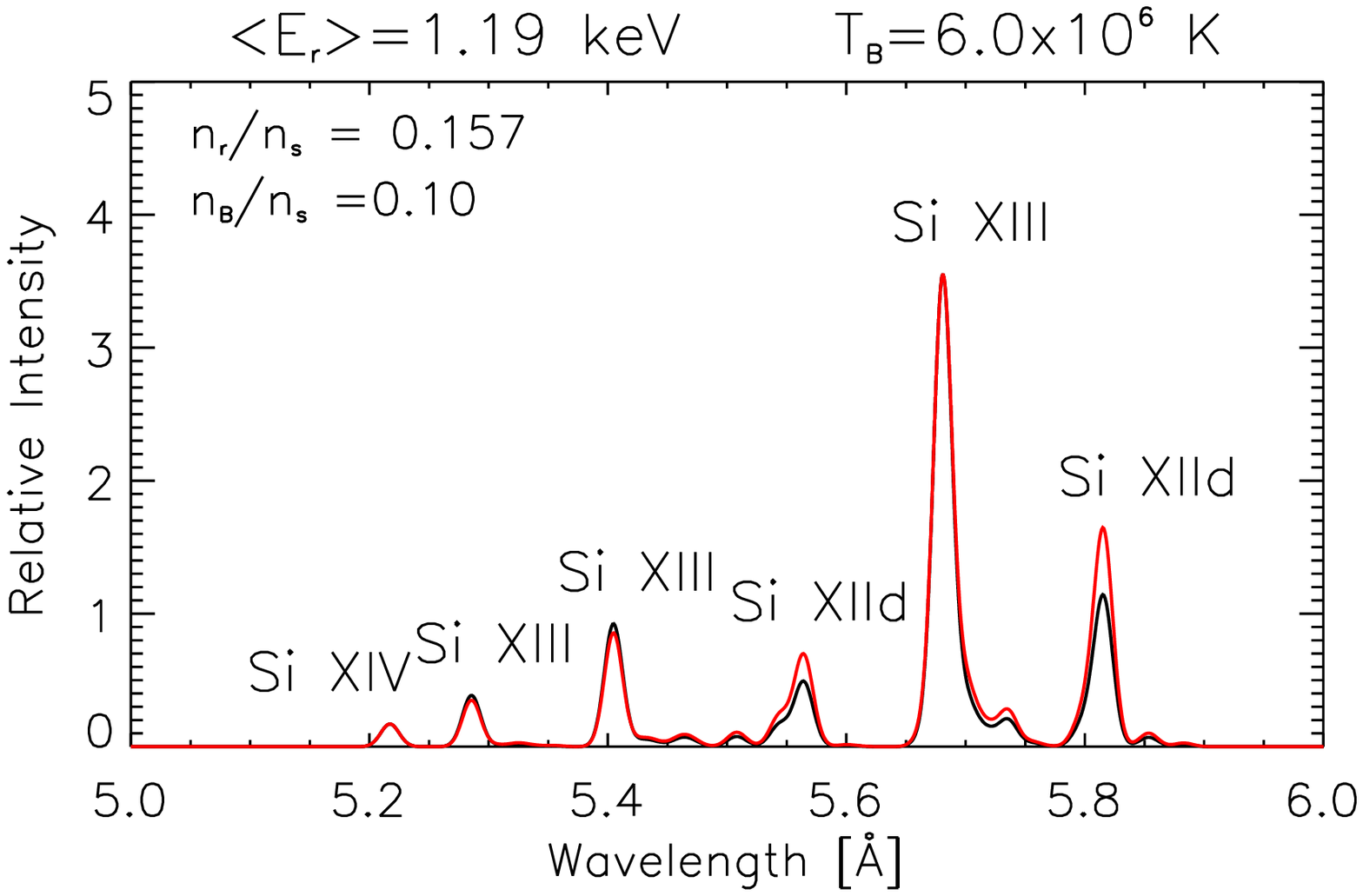}
\caption{Synthetic spectra of the composed distributions (red lines) in dependence on the background temperature
$T_B$ for low electron densities of background plasma (from top to bottom). For comparison, the spectra of Maxwellian distributions
are added (black lines).}
\label{sp_comp3}
\end{figure}

\begin{figure}
\centering
\includegraphics[width=8.4cm]{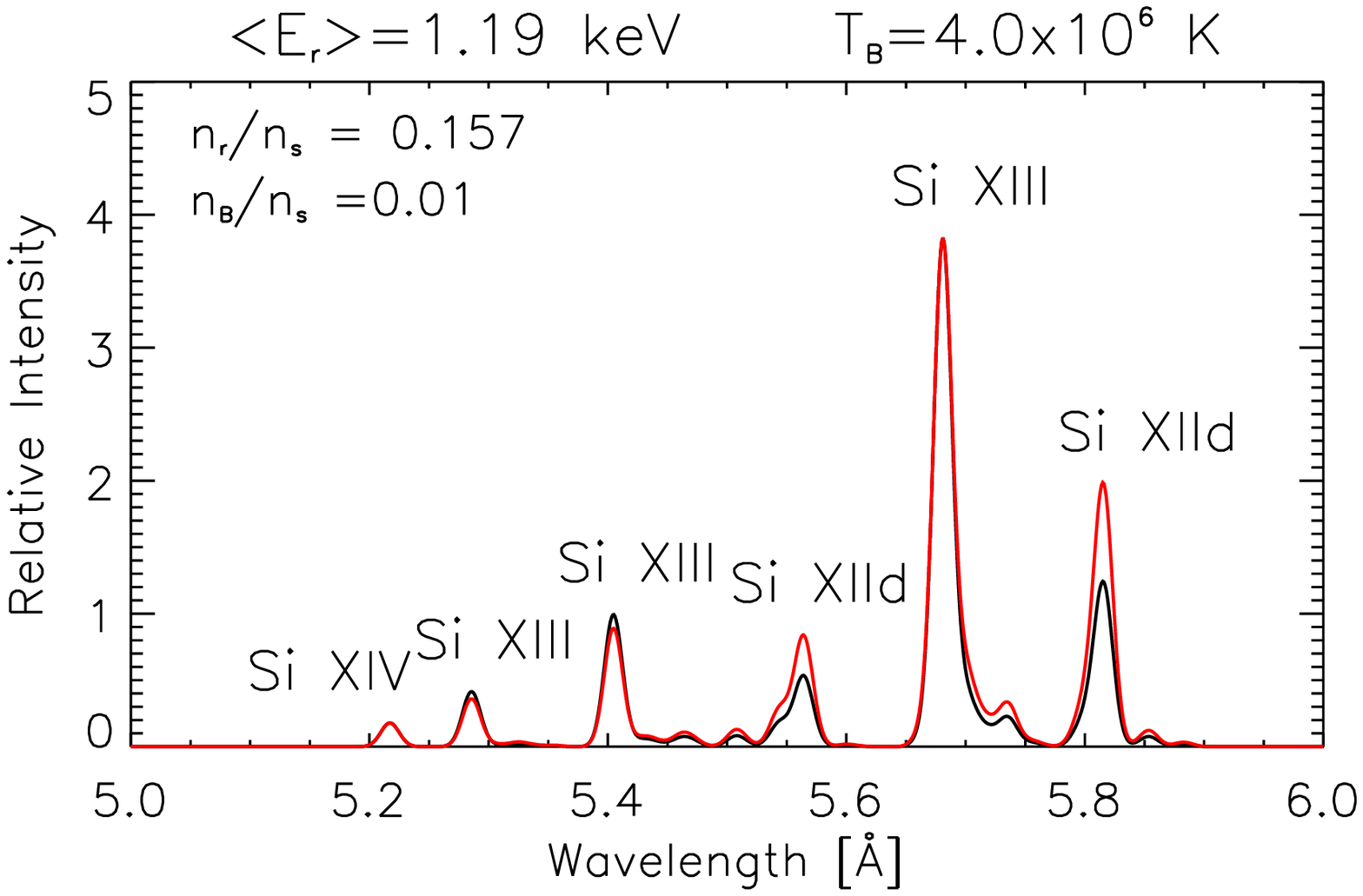}
\includegraphics[width=8.4cm]{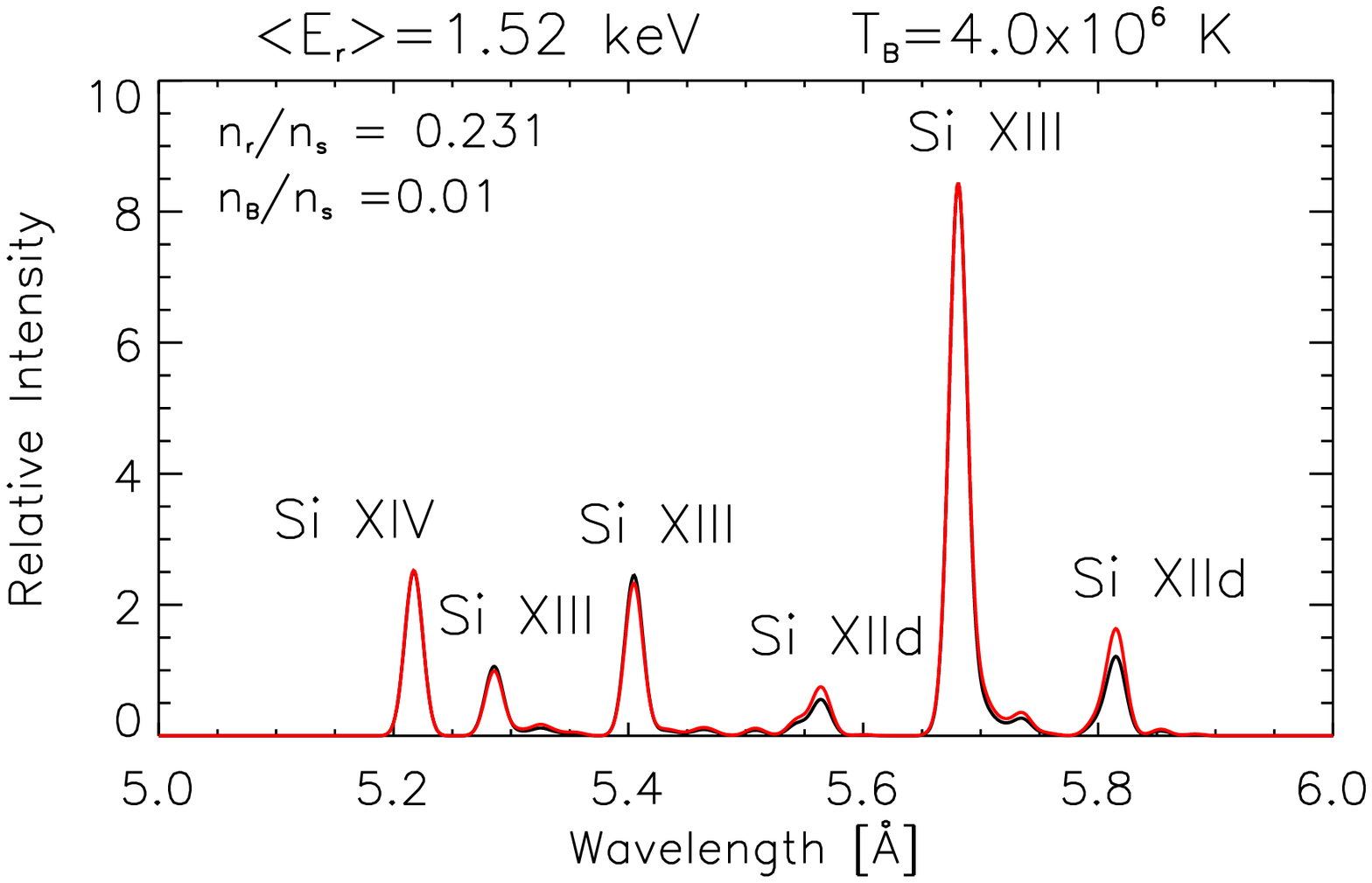}
\includegraphics[width=8.4cm]{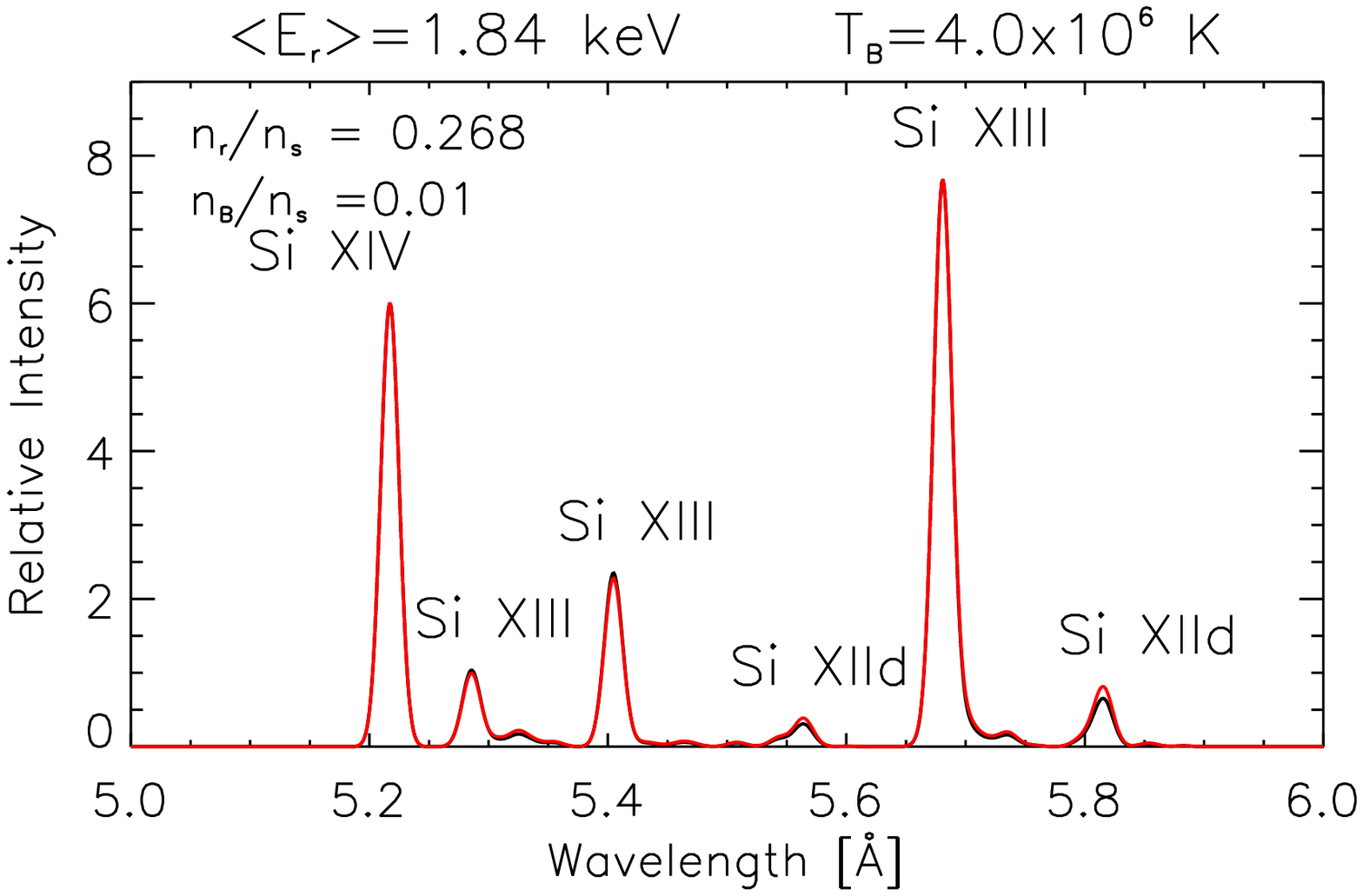}
\caption{Synthetic spectra of the composed distributions (red lines) in dependence on the mean energy
$\langle E_r \rangle$ of reflected electrons (or the temperature $T_s$
of the seed plasma) (from top to bottom). For comparison, the spectra of Maxwellian distributions
are added (black lines). }
\label{sp_comp2}
\end{figure}

\begin{figure}
\centering
\includegraphics[width=8.4cm]{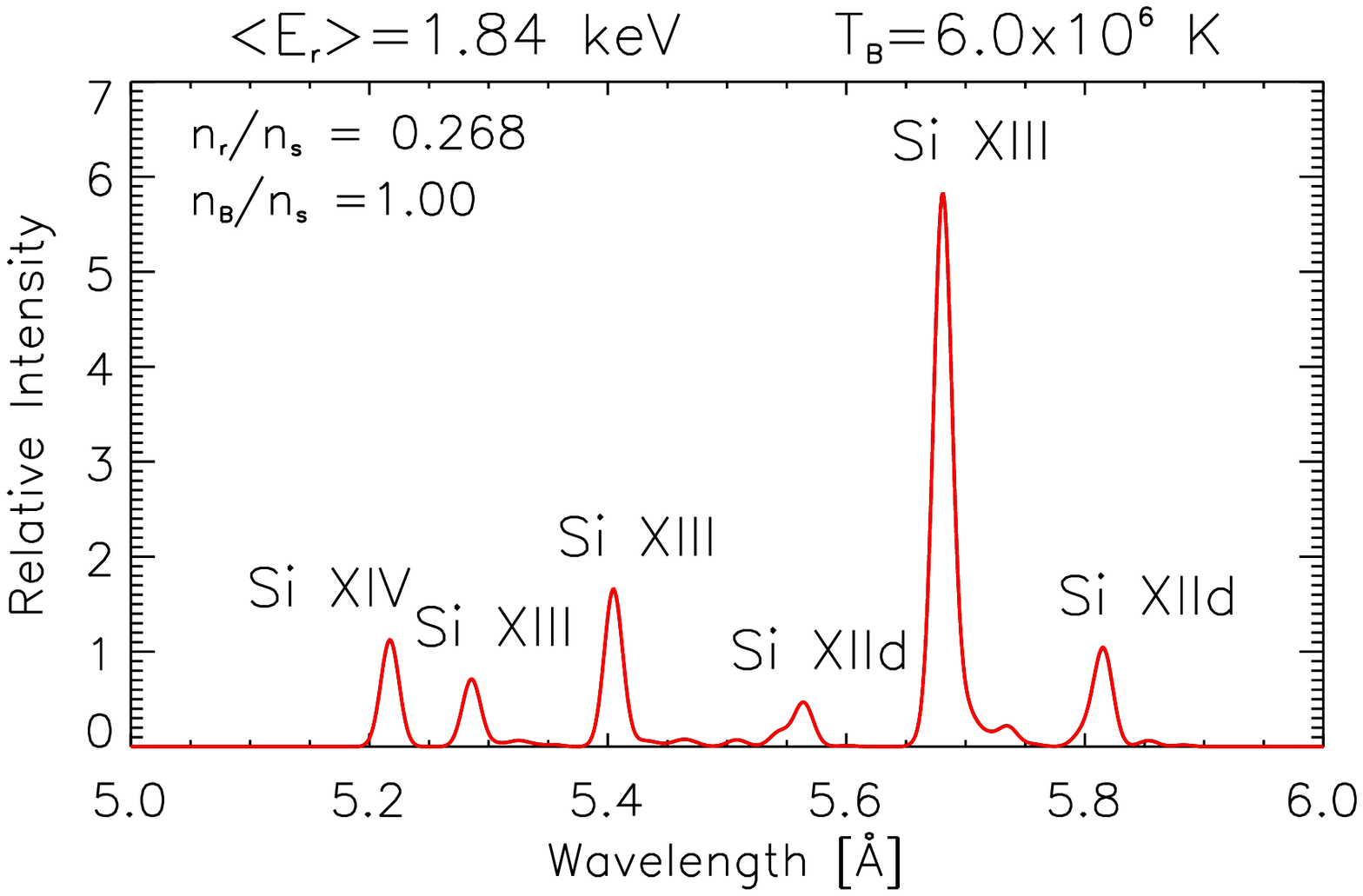}
\includegraphics[width=8.4cm]{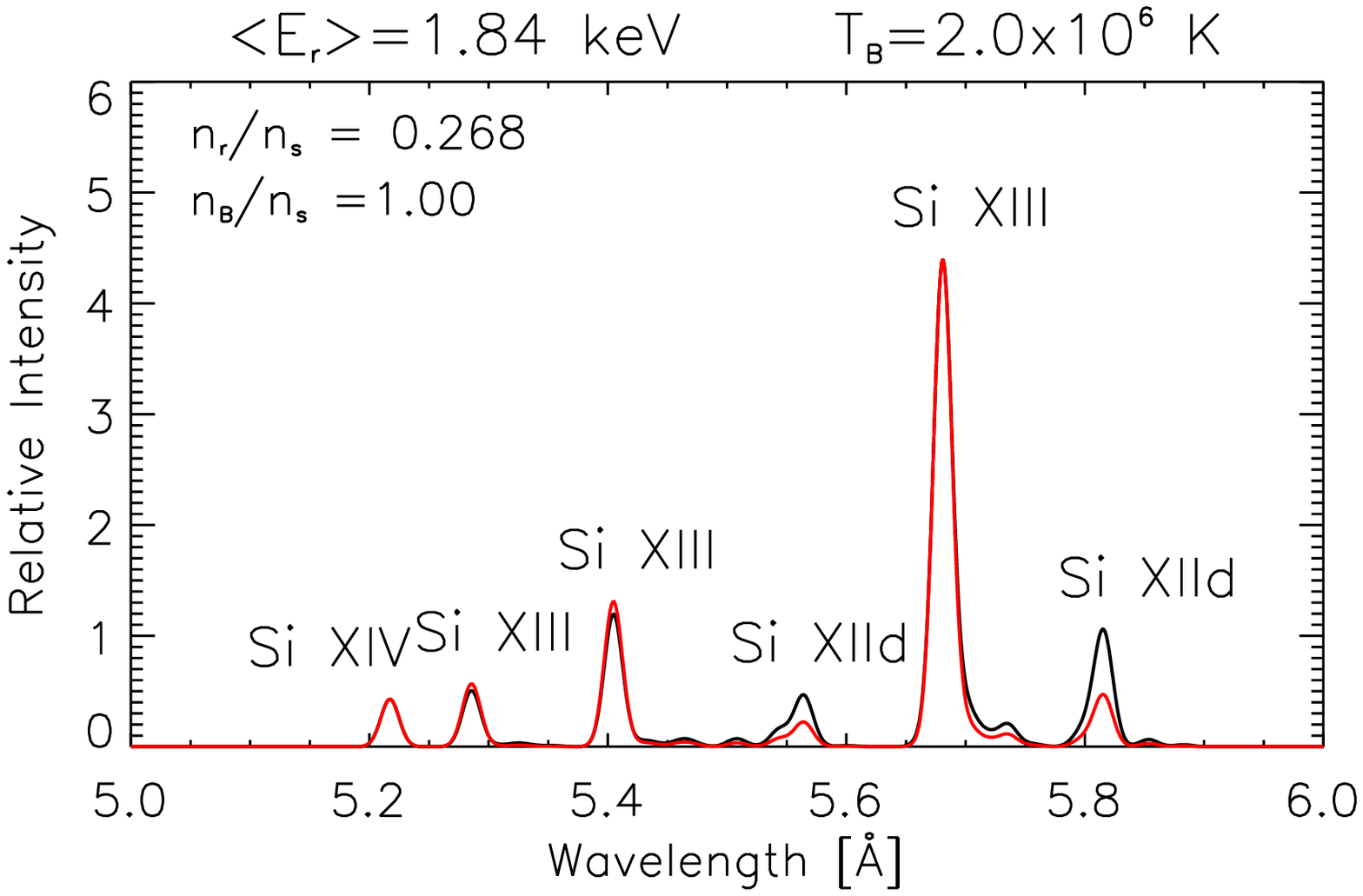}
\caption{Synthetic spectra of the composed distributions (red lines) in dependence on the background temperature
$T_B$ for high electron densities of background plasma (from top to bottom). For comparison, the spectra of Maxwellian distributions
are added (black lines).}
\label{sp_comp4}
\end{figure}

\section{X-ray line spectra for the shock-reflected electrons together with different
background electron distributions}

Considering different mean energies of the reflected electrons (cases~1--5, Tab.~\ref{tab1}) and the background plasma with different parameters we calculated the X-ray line spectra and compared them with Maxwellian ones.

We found that the spectra with the enhanced intensities of the \ion{Si}{xii}d satellite lines similar to the $n$-distribution spectra are formed for the low mean energies of the shock-reflected electrons. The enhancement is higher for the lower electron densities of the background plasma (Fig.~\ref{sp_comp1}). The top spectrum in Fig.~\ref{sp_comp1} is formed by the composed distribution, which is more similar to the $n$-distribution not only in its high-energy part but also at low energies, as can be seen by the comparison of the distribution for $n$\,=\,3  in Fig.~\ref{n_dst} with the composed distribution  for $\langle E_r\rangle$\,=\,1.68~keV and $n_B/n_s$=0.01 in Fig.~\ref{Fig:distr} (top). The low number of the low-energy electrons results in a decrease of the recombination rate and increase of the ionization degree, which causes higher relative intensity of \ion{Si}{xiv} 5.22 \AA\,in comparison with \ion{Si}{xiii} 5.68 \AA\, (Fig.~\ref{sp_comp1}, {top}).

The temperature of the background plasma has only a small effect on the spectra at the low background electron densities (Fig.~\ref{sp_comp3}) because the changes of the recombination rates with temperature $T_B$ are smaller than the changes with $n_B/n_s$. For the high densities of the background electrons, the Maxwellian part of the composed distribution becomes more important and spectra look more or less Maxwellian  (Fig.~\ref{sp_comp1}, bottom).

The most distinct effect of the shock-reflected electrons on the spectra was found in the case with the low background plasma densities (Fig.~\ref{sp_comp2}). With increase of the mean energy of reflected electrons, the relative intensity of \ion{Si}{xiv} line increases and the enhancement of the satellite line intensities decreases. The shift of the energy maximum of the shock-reflected electrons reflects an increase in the number of electrons capable of exciting and ionizing the \ion{Si}{xiii}  ions (Fig.~\ref{satellite}, hatched areas are much larger for distribution with $\langle E_r\rangle$\,=\,1.84 keV). This rise results in a strong increase of the ionization and excitation rates. However, the increase in the number of electrons with energy $E_d$ is not so strong (Fig.~\ref{satellite}), therefore the intensity of the satellite lines decreases with regard to the allowed lines.

The \ion{Si}{xii}d satellite line enhancement disappears for the high mean energies of the shock-reflected electrons and high electron densities of the
background plasma (Fig.~\ref{sp_comp4}, {top}). On the contrary, the opposite effect, that is, a decrease in the \ion{Si}{xii}d satellite
line intensities in comparison with the Maxwellian case, can be observed (Fig.~\ref{sp_comp4}, {bottom}). This decrease is higher for the low
temperatures of the background plasma. In this case the shock reflected electrons strongly influence total energy of the composed distribution and form
the high-energy tail of the distribution. The shape of these distributions is similar to the Maxwellian distribution with a power law tail and also to the
$\kappa$-distributions, and naturally forms similar spectra. However, the increase in the background temperature leads to an increase in  the number of
high-energy electrons in the Maxwellian distribution and inhibits the effect of the shock-reflected electrons on the composed distribution, and spectra  become Maxwellian (Fig.~\ref{sp_comp4}, top).

\section{Discussion}

Now a question arises concerning whether or not, in the impulsive phase of some solar flares, where the enhanced intensities of the \ion{Si}{xii}d satellite lines were observed, there are regions in some distance from shocks with much lower densities where the reflected electrons propagate and thus produce these intense satellite lines.

Besides shocks that are believed to be generated in connection with the magnetic field reconnection and plasmoid formation and their interactions \citep{2008A&A...477..649B,2016ApJ...827L...3V}, the so called termination shock is expected to be formed near the upper part of the flare arcade due to fast outflowing plasma from the reconnection site lying directly above \citep[see the flare scenario, e.g., Fig.~1 in the paper by][]{2009A&A...494..669M}.  This termination shock could be that where the electrons are accelerated and reflected. To fulfill conditions for the enhanced intensities of \ion{Si}{xii}d satellite lines, the reflected electrons need to propagate to  regions with lower
densities. Are there such regions in the flare cusp structure at some distance from the termination shock?

 During propagation of the shock-reflected electrons the bump-on-tail instability can generate Langmuir waves and then the radio emission. This process by itself is important in
detection of the flare termination shock~\citep{2015Sci...350.1238C}. However, we expect that the plasma in the flare cusp structure in front of the termination shock is in the state of a strong turbulence with some low-density regions. We note that the presence of the turbulence can reduce an effect of the bump-on-tail instability \citep[][page 210]{1980gbs..bookR....M}.  Nevertheless, at the present state of observations and modelling we cannot confirm these low-density regions. Another possibility is that the shock-reflected electrons propagate out of the flare cusp structure, where there are certainly such low-density regions.

Our results have some general implications. They are valid for any nearly perpendicular shocks generated in solar flares (standing or propagating) with an appropriate low-density region aside from these shocks.

 In order for the distribution of reflected electrons to be similar to an $n$-distribution function, their source distribution must be Maxwellian or very close to it. Moreover, the resulting distribution is formed by single reflections inside the shock layer in our calculations, so the role of random processes, such as pitch-angle scattering, is neglected. If some of these assumptions were not in fact valid, the distribution of reflected electrons would resemble a $\kappa$-distribution function rather than an $n$-distribution function \citep{1991BAICz..42...65V}.

Our model implicitly assumes processes in the collisionless regime, which gives an upper limit for plasma densities considered immediately in front of the shock of about $10^{11}$ cm$^{-3}$. Furthermore, the wave-particle interactions can also modify the distribution function of the reflected electrons propagating some distance from the shock to low-density regions. Nevertheless we assume that the energy distribution of these electrons is still appropriate to generate the enhanced intensities \ion{Si}{xii}d satellite lines. It can be supported by in-situ observations of reflected electrons from planetary bow shocks, which are detectable as beams very far from their source regions. \citet{A81} reports
observations of reflected electrons by Earth's bow shock from a distance of 1.5~Gm (nearly two solar radii).

\section{Conclusions}

 We have found that the high-energy part of the $n$-distribution function can be matched by a distribution function of reflected electrons at a nearly perpendicular shock. The effective $n$ depends on shock-wave parameters; it increases with decreasing magnetic field jump at the shock wave, decreasing cross-shock electrostatic potential, increasing of shock velocity, or with increasing $\theta_{Bn}$. Furthermore, it increases above 10 very rapidly with $\theta_{Bn}$ over $87^\circ$. Temporal variations of $n$ following from observations \citep{2011A&A...533A..81K} could be attributed to temporal changes of shock-wave parameters, or changes of a shock region where reflected electrons come from.

We have shown that the distribution of the shock-reflected electrons influences the relation between the excitation of the allowed and satellite lines and their relative intensities. Spectra similar to $n$-distribution spectra can be formed by the distribution composed from the shock-reflected electrons with low mean energy and background Maxwellian plasma at lower electron densities and temperatures. This combination of parameters enables the shock-reflected electrons to dominate the electron distribution shape and to enhance intensities of the \ion{Si}{xii}d satellite lines.  Conversely, the high electron densities of the background plasma inhibit the effect of the reflected electrons on the shape of the distribution function and spectra come to be Maxwellian. In an extreme case, for the low temperature and high density background plasma, the reflected electrons form the high-energy tail of the distribution, which has an opposite effect on the satellite line intensities compared to the $n$-distribution and resembles an effect of the $\kappa$-distribution function. It is interesting that the same mechanism, but under different conditions, can produce spectral effects attributed to the $n$- or $\kappa$-distribution functions.

Apparently, the formation of the spectrum with high intensities of the \ion{Si}{xii}d satellite lines by the shock-reflected electrons needs specific conditions. To decipher whether or not these conditions are really present in the impulsive phase of some solar flares, further and more detailed observations are required.

\begin{acknowledgements}
We acknowledge support from Grants P209/12/0103, 17-16447S, 14-19376S, 17-06065S, 15-17490S, and 16-13277S by the Grant Agency of the Czech Republic and from the
AV~\v{C}R grant RVO:67985815.
\end{acknowledgements}

\bibliographystyle{aa}
\bibliography{distr_r2}

\end{document}